\documentclass[twocolumn]{autart}    

\usepackage{graphicx}          

\usepackage{amsmath,amssymb,bm,mathtools}
\usepackage{epstopdf}
\usepackage{pgfplots}
\usepackage{enumitem}
\pgfplotsset{compat=1.18}
\usepackage{tikz}
\usetikzlibrary{decorations.pathreplacing,decorations.markings}
\newcommand*\circled[1]{\tikz[baseline=(char.base)]{
            \node[shape=circle,draw,inner sep=1pt] (char) {#1};}}
\DeclareMathOperator{\rank}{rank}
\DeclareMathOperator{\diag}{diag}
\DeclareMathOperator{\col}{col}
\newcommand{\tikzlot}[1]{%
	\draw[gray,fill=gray!10] (0,0) -- (2,0) -- (2,2) -- (0,0);
	\draw[latex-latex] (2.5,0) node[below] {$z$} -- (0,0) -- (0,2.5) node[left] {$\zeta$};
	\draw (0,2) node[left] {$1$};
	\draw (2,0) node[below] {$1$};
	\draw (0,0) node[left] {$0$};
	\draw (0,0) node[below] {$0$};
	\draw (1,2.2) node {#1};}

\newcommand{\tint}{\textstyle\int}
\newcommand{\both}{\diamond}

\usepackage[acronym]{glossaries}
\glsdisablehyper 
\newacronym[shortplural={BCs}, longplural={boundary conditions}]{BC}{BC}{boundary condition}
\newacronym[shortplural={ICs}, longplural={initial conditions}]{IC}{IC}{initial condition}

\begin{document}

\begin{frontmatter}

\title{Using dynamic extensions for the backstepping control of hyperbolic systems} 

\thanks[footnoteinfo]{This paper was not presented at any IFAC 
meeting. Corresponding author N.~Gehring.}

\author[jku]{N.~Gehring\thanksref{footnoteinfo}}\ead{nicole.gehring@jku.at}, ~
\author[ulm]{J.~Deutscher}\ead{joachim.deutscher@uni-ulm.de}, ~
\author[uds]{A.~Irscheid}\ead{a.irscheid@lsr.uni-saarland.de}

\address[jku]{Institute of Automatic Control and Control Systems Technology, Johannes Kepler University Linz, Linz, Austria}
\address[ulm]{Institute of Measurement, Control and Microtechnology, Ulm University, Ulm, Germany}
\address[uds]{Chair of Systems Theory and Control Engineering, Saarland University, Saarbr\"ucken, Germany}


\begin{abstract} 
This paper systematically introduces dynamic extensions for the boundary control of general heterodirectional hyperbolic PDE systems.
These extensions, which are well known in the finite-dimensional setting, constitute the dynamics of state feedback controllers.
They make it possible to achieve design goals beyond what can be accomplished by a static state feedback.
The design of dynamic state feedback controllers is divided into first introducing an appropriate dynamic extension and then determining a static feedback of the extended state, which includes the system and controller state, to meet some design objective.
In the paper, the dynamic extensions are chosen such that all transport velocities are homogenized on the unit spatial interval.
Based on the dynamically extended system, a backstepping transformation allows to easily find a static state feedback that assigns a general dynamics to the closed-loop system, with arbitrary in-domain couplings.
This new design flexibility is also used to determine a feedback that achieves complete input-output decoupling in the closed loop with ensured internal stability.
It is shown that the modularity of this dynamic feedback design allows for a straightforward transfer of all results to hyperbolic PDE-ODE systems.
An example demonstrates the new input-output decoupling approach by dynamic extension.
\end{abstract}

\end{frontmatter}

\section{Introduction}

It is well known from multivariable finite-dimensional systems that dynamic controllers are very useful in meeting a multitude of design objectives.
In a time-domain design using a state representation, dynamic extensions are of particular interest.
They are chosen so that the augmented system can be stabilized by a static state feedback of the extended state (see, e.g., \cite{Isidori1995} in the context of achieving full relative degree or \cite{Levine2009} concerning flatness-based control).
In addition to facilitating the stabilization task, dynamic extensions are used, for example, to ensure the input-output decoupling of linear finite-dimensional systems (see \cite{Morse1970}) and dynamic state feedback linearization for nonlinear systems (see \cite{Charlet1991}).
Recently, dynamic extensions also turned out to be useful in ODE backstepping (see \cite{Reger2019CDC}).
This naturally raises the question of how to introduce (infinite-dimensional) dynamic extensions in the backstepping framework for PDEs, as well as the control of PDEs in general, and what additional design flexibilities can be obtained.
While dynamic controllers for PDEs are usually found in the context of observer-based feedback designs (see, e.g., \cite{Krstic2008book,Curtain2020}), which are of no further interest here, results beyond this are very scarce, to the best of the authors' knowledge.
One example is the flatness-based design of a dynamic state feedback for a drilling system in \cite{Knueppel_2014_chapter}.
A breakthrough was made by the design in \cite{Redaud2022CDC,Redaud2024SCL}, where the well-known backstepping control of PDEs is combined with the port-Hamiltonian framework for the boundary stabilization of hyperbolic PDEs.
Importantly, the approach allows the choice of target systems, i.e., closed-loop dynamics, with arbitrary in-domain couplings.
While this seems to contradict established results on the solvability of the kernel equations associated with the backstepping transformation, where additional local terms have to be introduced in the target system to ensure well-posedness in the multivariable case (see \cite{Hu2019siam,Hu2016tac}), it does in fact not.
Contrary to the design of static state feedbacks in \cite{Hu2019siam,Hu2016tac}, the new time-affine transformation in \cite{Redaud2022CDC,Redaud2024SCL} implicitly introduces a controller dynamics.

This paper systematically presents dynamic extensions for heterodirectional hyperbolic PDE systems with spatially-varying transport velocities based on an alternative interpretation of the results in \cite{Redaud2022CDC,Redaud2024SCL} for systems with constant transport velocities.
The new systematic and modular design opens up previously untapped potential in the control design, both w.r.t.\ system classes that might be considered and control objectives that can be met.
The concept of dynamic extensions is presented in the context of homogenizing the transport velocities.
The idea is to introduce a dynamic extension in a manner that ensures all state components propagating from left to right and right to left, respectively, exhibit identical transport velocities and, thus, delays in the extended system.
Hence, the dynamic state feedback controller can change the transport velocities of the dynamically extended system and the closed loop, unlike static state feedback controllers.
Based on the dynamically extended system, similar to \cite{Redaud2022CDC,Redaud2024SCL}, the backstepping design of a static feedback of the extended state, i.e., the system and the controller state, using only on well-known kernel equations is straightforward and allows the choice of general target systems with arbitrary in-domain couplings.
Despite the dynamic extension, which artificially delays the control input, the convergence to zero in minimum time is preserved (cf.\ static state feedback in \cite{Auriol2016AUT}), provided that the design parameters are chosen appropriately.
In addition, dynamic controllers are designed such that the components of the input and some given output are completely decoupled in the closed-loop transfer behavior (see also \cite{Curtain1985,Otsuko1990}).
Thanks to the modularity of the dynamic feedback design, the concept of dynamic extensions as well as the assignment of general closed-loop dynamics and the decoupling of inputs and outputs are also easily transferred to hyperbolic PDE-ODE systems.

The paper is organized as follows.
First, Section~\ref{sec:problem} presents the problem formulation and the basic design idea.
In Section~\ref{sec:homogenization}, a dynamic extension is introduced in order to homogenize the transport velocities.
Based on the dynamically extended system, static feedbacks of the extended state are derived in Section~\ref{sec:applications} to meet different design objectives.
This includes the treatment of PDE-ODE systems.
The usefulness of the new controller design is illustrated in Section~\ref{sec:example} by achieving a decoupled input-output behavior for a dynamically extended closed-loop system.

\textit{Notation:}
Define
\begin{equation}
\label{eq:def_E}
    E_- = \begin{bmatrix} I_{n_-} \\ 0 \end{bmatrix} \in \Rset^{n\times n_-}, \quad E_+ = \begin{bmatrix} 0 \\ I_{n_+} \end{bmatrix} \in \Rset^{n\times n_+}
\end{equation}
for $n=n_-+n_+$, with identity matrices $I_r\in\Rset^{r\times r}$.
The elements of a vector $v\in\Rset^n$ are denoted by $v_i=e_i^\top v$, $i=1,\dots,n$, where $e_i$ is always the canonical unit vector of appropriate dimension.
Similarly, $m_{ij}$ denotes the element of a matrix $M=[m_{ij}]$ in the $i$-th row and $j$-th column.
The operator $\col$ stacks up its arguments, e.g., $M=\col(E_-^\top M,E_+^\top M)$. 
The notation $[M]_\ast$ means that only those elements of a matrix $M$ are considered that satisfy the condition $\ast$.
The use of $\both$ as a sub- or superscript indicates that a statement applies to both cases, $-$ and $+$.
For example, $\lambda_{n_\both}^\both$ refers to $\lambda_{n_-}^-$ and $\lambda_{n_+}^+$.

\section{Problem statement}
\label{sec:problem}

The general heterodirectional hyperbolic system
\begin{subequations}
\label{eq:sys}
    \begin{align}
    \label{eq:sys_pde}
        \partial_t x(z,t) &= \Lambda(z) \partial_z x(z,t) + A(z) x(z,t) \\
    \label{eq:sys_bc0}
        x_+(0,t) &= Q_0 x_-(0,t) \\
    \label{eq:sys_bc1}
        x_-(1,t) &= Q_1 x_+(1,t) + u(t)
    \end{align}
\end{subequations}
is defined for $(z,t)\in[0,1]\times\Rset^+$ with state $x(z,t)=\col(x_-(z,t),x_+(z,t))\in\Rset^n$, $n=n_-+n_+$. 
Due to the structure of
\begin{equation}
    \Lambda = \begin{bmatrix} \Lambda^- & 0 \\ 0 & -\Lambda^+ \end{bmatrix}, \qquad \Lambda^\both = \diag(\lambda_1^\both,\dots,\lambda_{n_\both}^\both)
\end{equation}
with elements $\lambda_i^\both\in C^1([0,1])$, $i=1,\dots,n_\both$, ordered according to
\begin{equation}
\label{eq:lambdas}
    \lambda_1^\both(z) > \cdots > \lambda_{n_\both}^\both(z) > 0, \qquad z\in[0,1],
\end{equation}
the $n_-$ components of $x_-(z,t)\in\Rset^{n_-}$ describe a transport in the negative $z$-direction, whereas the $n_+$ components of $x_+(z,t)\in\Rset^{n_+}$ propagate in the opposite direction.
Both transport processes are interconnected via $Q_0\in\Rset^{n_+\times n_-}$, $Q_1\in\Rset^{n_-\times n_+}$ and the coupling matrix $A(z)\in\Rset^{n\times n}$, with entries $a_{ij}\in C([0,1])$ as well as $a_{ii}(z)=0$ without loss of generality (see \cite{Hu2016tac}).
By the input $u(t)\in\Rset^{n_-}$, the system \eqref{eq:sys} is fully boundary actuated at $z=1$.
The \gls{IC} of \eqref{eq:sys} is $x(z,0)=x_0(z)$.

In the following, dynamic state feedback controllers are systematically designed for \eqref{eq:sys} using the backstepping approach.
For that, the design is split into multiple steps:
\begin{enumerate}[label=\protect\circled{\arabic*}]

    \item A dynamic extension is introduced such that the transport velocities of the dynamically extended system are homogenized, i.e., identical, for all components propagating in the positive and negative direction, respectively.

    \item The design of a static feedback of the extended state (comprising both the system and the controller state) can readily accommodate a variety of design specifications. Contrary to a static state feedback of the original system state $x(z,t)$, this allows for the assignment of general closed-loop dynamics with arbitrary in-domain couplings and a straightforward decoupling control, e.g., w.r.t.\ the output $x_-(0,t)$.

    \item The dynamic state feedback results from combining the dynamic extension, which constitutes the controller dynamics, with the static feedback of the extended state.
    
\end{enumerate}

The modularity of this approach also enables the straightforward design of dynamic state feedback controllers for PDE-ODE systems, where an additional ODE dynamics is considered at the unactuated boundary of \eqref{eq:sys} (see Section~\ref{sec:pdeode}).

\begin{rem}
    The results of this contribution apply in a straightforward way to systems with $\lambda_1^\both(z) \ge \cdots \ge \lambda_{n_\both}^\both(z) > 0$, $z\in[0,1]$, instead of \eqref{eq:lambdas} that could contain additional local terms $x_-(0,t)$ and Volterra integral terms in \eqref{eq:sys_pde} as well as additional terms in \eqref{eq:sys_bc1}.
    \hfill $\triangle$
\end{rem}

\section{Dynamic extension for homogenizing transport velocities}
\label{sec:homogenization}

In order to simplify the introduction of a dynamic extension that homogenizes the transport velocities of \eqref{eq:sys}, first, a preliminary backstepping transformation is used to eliminate the in-domain coupling due to $A(z)$, thus, essentially mapping \eqref{eq:sys} into cascades of transport equations.
A scaling of the spatial domain of each of the transport equations then allows to homogenize the propagation speeds and to define an appropriate dynamic extension that return all transport processes to the same spatial interval, which ensures identical transport delays between the two boundaries.
These new ideas are motivated in Section~\ref{sec:motex} by means of a simple example.
The final result is given in Section~\ref{sec:ext_result}.

\subsection{Preliminary backstepping transformation}
\label{sec:trafo1}

Consider the Volterra integral transformation
\begin{equation}
\label{eq:trafo1}
    \tilde x(z,t) = x(z,t) - \tint_0^z K(z,\zeta) x(\zeta,t)\d\zeta
\end{equation}
with the kernel $K(z,\zeta)\in\Rset^{n\times n}$ on the triangular domain
\begin{equation}
    \mathcal T = \{(z,\zeta)\in[0,1]^2|\zeta\le z\}
\end{equation}
in order to map system \eqref{eq:sys} into a form
\begin{subequations}
\label{eq:sys_intermediate}
    \begin{align}
    \label{eq:sys_intermediate_pde}
        \partial_t \tilde x(z,t) &= \Lambda(z) \partial_z \tilde x(z,t) + A_0(z) \tilde x_-(0,t) \\
    \label{eq:sys_intermediate_bc0}
        \tilde x_+(0,t) &= Q_0 \tilde x_-(0,t) \\
    \label{eq:sys_intermediate_bc1}
        \tilde x_-(1,t) &= \tilde u(t)
    \end{align}
\end{subequations}
without in-domain couplings between the transport PDEs, where the new input
\begin{equation}
\label{eq:utilde}
    \tilde u(t) = Q_1x_+(1,t) + u(t) - \tint_0^1 E_-^\top K(1,\zeta)x(\zeta,t)\d\zeta
\end{equation}
is introduced for ease of notation.
For that, straightforward calculations reveal that $K(z,\zeta)$ has to satisfy the well-known kernel equations
\begin{subequations}
\label{eq:kernel1}
    \begin{align}
    \label{eq:kernel1_pde}
        \Lambda(z)\partial_z K(z,\zeta) + \partial_\zeta(K(z,\zeta)\Lambda(\zeta)) &= K(z,\zeta)A(\zeta) \\
    \label{eq:kernel1_zz}
        K(z,z)\Lambda(z) - \Lambda(z)K(z,z) &= A(z) \\
    \label{eq:kernel1_z0}
        [K(z,0)\Lambda(0)(E_-+E_+Q_0)]_{i\le j} &= 0
    \end{align}
\end{subequations}
on $\mathcal T$.
The existence of a unique piecewise $C(\mathcal T)$-solution $K(z,\zeta)$ of \eqref{eq:kernel1} is verified in \cite{Hu2019siam} based on appropriate additional \glspl{BC} for the kernel at the boundary $z=1$ that can be freely chosen.
The elements of $A_0=\col(A_0^-,A_0^+)$, partitioned into a strictly lower triangular matrix $A_0^-(z)\in\Rset^{n_-\times n_-}$ and a matrix $A_0^+(z)\in\Rset^{n_+\times n_-}$ without a special form are defined based on the solution $K(z,\zeta)$:
\begin{equation}
    [A_0(z)]_{i>j} = [K(z,0)\Lambda(0)(E_-+E_+Q_0)]_{i>j}.
\end{equation}

\subsection{Idea of homogenization}
\label{sec:motex}

In order to illustrate the idea of homogenization, consider a system of the form \eqref{eq:sys_intermediate} with $n_-=n_+=2$ and constant transport velocities:
\begin{subequations}
\label{eq:motiv_sys}
    \begin{align}
    \label{eq:motiv_sys_m}
        \partial_t \tilde x_-(z,t) &= \!\begin{bmatrix} \lambda_1^- & 0 \\ 0 & \lambda_2^- \end{bmatrix} \partial_z \tilde x_-(z,t) + \!\begin{bmatrix} 0 & 0 \\ a_0^-(z) & 0 \end{bmatrix} \tilde x_-(0,t) \\
    \label{eq:motiv_sys_p}
        \partial_t \tilde x_+(z,t) &= \!\begin{bmatrix} -\lambda_1^+ & 0 \\ 0 & -\lambda_2^+ \end{bmatrix} \partial_z \tilde x_+(z,t) + A_0^+(z) \tilde x_-(0,t) \\
    \label{eq:motiv_sys_0}
        \tilde x_+(0,t) &= Q_0 \tilde x_-(0,t) \\
    \label{eq:motiv_sys_1}
        \tilde x_-(1,t) &= \tilde u(t).
    \end{align}
\end{subequations}
Therein, the presence of the terms involving $a_0^-(z)$ and $A_0^+(z)$ is a direct consequence of the backstepping transformation \eqref{eq:trafo1}.
It should be noted that they cannot be chosen freely.
However, if $\lambda_1^-=\lambda_2^-$, as indicated in \cite[Rem.~6]{Hu2016tac}, the choice of $a_0^-(z)$ and $A_0^+(z)$ is arbitrary and, in particular, one can set both equal to zero.

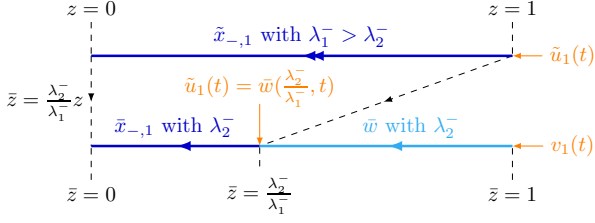
\begin{figure}
    \centering
    \scalebox{.8}{\begin{tikzpicture}
    \tikzset{%
        mid arrow/.style={%
            postaction={decorate,decoration={markings,mark=at position .51 with {\arrow{latex}}}}},
        mid arrow reversed/.style={%
            postaction={decorate,decoration={markings,mark=at position .51 with {\arrowreversed{latex}}}}},
        mid arrow reversed double/.style={%
            postaction={decorate,decoration={markings,mark=at position .52 with {\arrowreversed{latex}},mark=at position .5 with {\arrowreversed{latex}}}}}
    }
    \definecolor{colorInput}{RGB}{253,127,0}
    \definecolor{colorSys}{RGB}{0,0,200}
    \definecolor{colorExt}{RGB}{51,171,241}
    \pgfmathsetmacro{\posnull}{3}
    \pgfmathsetmacro{\posmid}{5.8}
    \pgfmathsetmacro{\posend}{10}
    \pgfmathsetmacro{\posu}{11}
    \pgfmathsetmacro{\posz}{2.8}
    \pgfmathsetmacro{\poseins}{2}
    \pgfmathsetmacro{\poseinsbar}{0.5}
    \pgfmathsetmacro{\poszbar}{-.3}

    \node (z0) at (\posnull,\posz) {$z=0$} ;
    \node (z1) at (\posend,\posz) {$z=1$} ;
    \node (zbar0) at (\posnull,\poszbar) {$\bar z=0$} ;
    \node (zbarmid) at (\posmid,\poszbar) {$\bar z=\frac{\lambda_2^-}{\lambda_1^-}$} ;
    \node (zbar1) at (\posend,\poszbar) {$\bar z=1$} ;
    \draw[dashed] (zbar0) -- (\posnull,\poseinsbar) ;
    \draw[dashed] (zbarmid) -- (\posmid,\poseinsbar) ;
    \draw[dashed] (zbar1) -- (\posend,\poseinsbar) ;
    \draw[dashed] (z0) -- (\posnull,\poseins) ;
    \draw[dashed] (z1) -- (\posend,\poseins) ;

    \draw[mid arrow, dashed] (\posnull,\poseins) -- node[midway, left] {$\bar z=\frac{\lambda_2^-}{\lambda_1^-}z$} (\posnull,\poseinsbar) ;
    \draw[mid arrow, dashed] (\posend,\poseins) -- (\posmid,\poseinsbar) ;

    \draw[mid arrow reversed, very thick, colorSys] (\posnull,\poseinsbar) -- node[midway, above] {$\bar x_{-,1}$ with $\lambda_2^-$} (\posmid,\poseinsbar) ;
    \draw[mid arrow reversed double, very thick, colorSys] (\posnull,\poseins) -- node[midway, above] {$\tilde x_{-,1}$ with $\lambda_1^->\lambda_2^-$} (\posend,\poseins) ;

    \draw[mid arrow reversed, very thick, colorExt] (\posmid,\poseinsbar) -- node[pos=.6, above] {$\bar w$ with $\lambda_2^-$} (\posend,\poseinsbar) ;

    \node (ubar1) at (\posu,\poseinsbar) {\color{colorInput}{$v_1(t)$}} ;
    \draw[latex-, colorInput] (\posend,\poseinsbar) -- (ubar1) ;
    \node (u1) at (\posu,\poseins) {\color{colorInput}{$\tilde u_1(t)$}} ;
    \draw[latex-, colorInput] (\posend,\poseins) -- (u1) ;
    \draw[latex-, colorInput] (\posmid,\poseinsbar) -- node[pos=.75, above] {$\tilde u_1(t)=\bar w(\frac{\lambda_2^-}{\lambda_1^-},t)$} (\posmid,1.2) ;
\end{tikzpicture}}
    \caption{Idea of homogenization by dynamic extension.}
    \label{fig:homogen}
\end{figure}

This motivates the idea of a dynamic extension in the case where $\lambda_1^->\lambda_2^-$ in order to homogenize the transport velocities.
First, introduce $\bar z=(\lambda_2^-/\lambda_1^-)z$ and $\bar x_{-,1}(\bar z,t)=\tilde x_{-,1}(z,t)$ to equivalently represent the transport for the first component of $\tilde x_-(z,t)$ in \eqref{eq:motiv_sys_m} by
\begin{subequations}
    \begin{equation}
    \label{eq:motiv_sys_trans_m1}
        \partial_t \bar x_{-,1}(\bar z,t) = \lambda_2^- \partial_{\bar z} \bar x_{-,1}(\bar z,t), \qquad \bar z\in[0,\tfrac{\lambda_2^-}{\lambda_1^-}),
    \end{equation}
    i.e., by a slower transport on a smaller domain (see Figure~\ref{fig:homogen}), with the corresponding \gls{BC}
    \begin{equation}
    \label{eq:motiv_sys_trans_bc1}
        \bar x_{-,1}(\tfrac{\lambda_2^-}{\lambda_1^-},t) = \tilde u_1(t)
    \end{equation}
\end{subequations}
instead of the first line of \eqref{eq:motiv_sys_1}.
If the dynamic extension%
\begin{subequations}
\label{eq:motiv_ext}
    \begin{align}
    \label{eq:motiv_ext_pde}
        \partial_t \bar w(\bar z,t) &= \lambda_2^- \partial_{\bar z} \bar w(\bar z,t), \qquad \bar z\in[\tfrac{\lambda_2^-}{\lambda_1^-},1) \\
    \label{eq:motiv_ext_bc}
        \bar w(1,t) &= v_1(t) \\
    \label{eq:motiv_ext_output}
        \tilde u_1(t) &= \bar w(\tfrac{\lambda_2^-}{\lambda_1^-},t)
    \end{align}
\end{subequations}
as part of the controller is chosen to complement the domain of \eqref{eq:motiv_sys_trans_m1}, then $\bar x_{-,1}(\bar z,t)$ together with $\bar w(\bar z,t)$ as well as $\tilde x_{-,2}(z,t)$ not only share the same transport velocity $\lambda_2^-$ but also the same spatial domain $[0,1]$.
As sketched in Figure~\ref{fig:homogen}, the output \eqref{eq:motiv_ext_output} of the transport equation \eqref{eq:motiv_ext_pde} defines $\tilde u_1(t)$ in \eqref{eq:motiv_sys_trans_bc1}.
By the special choice of \eqref{eq:motiv_ext_pde}, 
\eqref{eq:motiv_ext} simply introduces a delayed new input $v_1(t)$, though a more complex dynamics with additional terms could be chosen instead of \eqref{eq:motiv_ext_pde}.
Thus, the control inputs $v_1(t)$ and $\tilde u_2(t)$ act synchronously on the boundary at $z=0$ with exactly the same delay, just like in the case $\lambda_1^-=\lambda_2^-$.

Introducing the composite state
\begin{subequations}
    \begin{align}
        \chi_{-,1}(z,t) &=
        \begin{cases}
            \bar x_{-,1}(z,t), & z\in[0,\tfrac{\lambda_2^-}{\lambda_1^-}] \\
            \bar w(z,t), & z\in(\tfrac{\lambda_2^-}{\lambda_1^-},1]
        \end{cases} \\
        \chi_{-,2}(z,t) &= \tilde x_{-,2}(z,t), \hspace{.6cm} z\in[0,1],
    \end{align}
\end{subequations}
where $z$ is used instead of $\bar z$, system \eqref{eq:motiv_sys} together with \eqref{eq:motiv_ext} can be written as the dynamically extended system%
\begin{subequations}
\label{eq:motiv_sys_ext}
    \begin{align}
    \label{eq:motiv_sys_ext_m}
        \partial_t \chi_-(z,t) &= \lambda_2^- \partial_z \chi_-(z,t) + \begin{bmatrix} 0 & 0 \\ \bar a_0^-(z) & 0 \end{bmatrix} \chi_-(0,t) \\
    \label{eq:motiv_sys_ext_p}
        \partial_t \tilde x_+(z,t) &= \begin{bmatrix} -\lambda_1^+ & 0 \\ 0 & -\lambda_2^+ \end{bmatrix} \partial_z \tilde x_+(z,t) + A_0^+(z) \chi_-(0,t) \\
    \label{eq:motiv_sys_ext_0}
        \tilde x_+(0,t) &= Q_0 \chi_-(0,t) \\
    \label{eq:motiv_sys_ext_1}
        \chi_-(1,t) &= \begin{bmatrix} v_1(t) \\ \tilde u_2(t) \end{bmatrix},
    \end{align}
\end{subequations}
with the piecewise definition of
\begin{equation}
    \bar a_0^-(z) =
    \begin{cases}
        a_0^-(\tfrac{\lambda_1^-}{\lambda_2^-}z,t), & z\in[0,\tfrac{\lambda_2^-}{\lambda_1^-}] \\
        0, & z\in(\tfrac{\lambda_2^-}{\lambda_1^-},1].
    \end{cases}
\end{equation}
Notably, the components of $\chi_-(z,t)$ share the same transport velocity, as was the intention of the homogenization.
Consequently, use of a classical backstepping transformation allows to map \eqref{eq:motiv_sys_ext_m} into a form without a local term, i.e., two transport equations that are not cascaded.
Appropriate choice of the inputs $v_1(t)$ and $\tilde u_2(t)$ in \eqref{eq:motiv_sys_ext_1} then results in a dynamical state feedback with the controller dynamics \eqref{eq:motiv_ext}.

The removal of all local terms in \eqref{eq:motiv_sys} is possible by homogenizing both the positive and negative transport velocities of the system.
While \eqref{eq:motiv_ext} introduces a preceding transport equation, thus artificially delaying the control action of the first input, homogenization in the positive $z$-direction is achieved by introducing a trailing transport equation for the component $\tilde x_{+,1}(z,t)$, which has the faster velocity $\lambda_1^+>\lambda_2^+$, at the end of the existing transport process.
This effectively delays an output $\tilde x_{+,1}(1,t)$.
The trailing transport equation is also part of the controller dynamics.

\subsection{Derivation of the dynamic extension}
\label{sec:ext_derivation}

In order to formulate the approach taken in the previous section for \eqref{eq:sys}, define the strictly increasing functions
\begin{equation}
    \phi_i^\both(z) = \tint_0^z \d \zeta/\lambda_i^{\both}(\zeta), \qquad i = 1,\dots,n_\both
\end{equation}
that represent transport times and denote by $\psi_i^\both(z)$ the corresponding inverse such that $\psi_i^\both(\phi_i^\both(z))=z$.
In order to homogenize the transport velocities in \eqref{eq:sys_intermediate}, based on $\lambda_{n_-}^-(z)$ and $\lambda_{n_+}^+(z)$ being the respective slowest propagation speeds (see \eqref{eq:lambdas}), the spatial variable $z$ is transformed for each of the PDEs in \eqref{eq:sys_intermediate_pde} according to
\begin{equation}
\label{eq:ztrafo}
    \bar z = \psi_{n_\both}^\both(\phi_i^\both(z)) \eqqcolon \sigma_i^\both(z), \quad i = 1,\dots,n_\both,
\end{equation}
mapping $[0,1]$ to a domain smaller or equal to the original one as $\sigma_i^\both(1)\le1$.
Note that $\bar z=z$ for $i=n_\both$, yet the index is included in \eqref{eq:ztrafo} to make the notation more compact.
Also, with a slight abuse of notation, the spatial variable $\bar z\in[0,\sigma_i^\both(1)]$ is used for all PDEs, even though each transport equation is on a different spatial domain following \eqref{eq:ztrafo}.
Then, introducing the transformed state components
\begin{equation}
\label{eq:xbar}
    \bar x_{\both,i}(\bar z,t) = \bar x_{\both,i}(\sigma_i^\both(z),t) = \tilde x_{\both,i}(z,t), \quad i = 1,\dots,n_\both
\end{equation}
with
\begin{align}
    \d_z \bar x_{\both,i}(\bar z,t) &= \partial_{\bar z} \bar x_{\both,i}(\bar z,t) \d_z\sigma_i^\both(z) \nonumber \\
    &= \partial_{\bar z} \bar x_{\both,i}(\bar z,t)\frac{\lambda_{n_\both}^\both(\bar z)}{\lambda_{i}^{\both}(z)} = \partial_z \tilde x_{\both,i}(z,t),
\end{align}
the PDEs in \eqref{eq:sys_intermediate_pde} for the components \eqref{eq:xbar} take the form
\begin{subequations}
\label{eq:sys_scaled_pde}
    \begin{align}
        \partial_t \bar x_{-,i}(\bar z,t) &= \phantom{-}\lambda_{n_-}^-(\bar z) \partial_{\bar z} \bar x_{-,i}(\bar z,t) + a_{-,i}^\top(\bar z) \bar x_-(0,t), \notag\\
        &\hspace{1cm} \bar z\in[0,\sigma_i^-(1)),\,\,\,\, i = 1,\dots,n_- \\
        \partial_t \bar x_{+,i}(\bar z,t) &= -\lambda_{n_+}^+(\bar z) \partial_{\bar z} \bar x_{+,i}(\bar z,t) +a_{+,i}^\top(\bar z) \bar x_-(0,t), \notag\\
        &\hspace{1cm} \bar z\in(0,\sigma_i^+(1)],\,\,\,\, i = 1,\dots,n_+,
    \end{align}
    where $a_{\both,i}^\top(\bar z) =  e_i^\top A_0^\both(\tau_i^\both(\bar z))$ and $z=\tau_i^\both(\bar z)$ is the inverse map of \eqref{eq:ztrafo}, i.e., $\tau_i^\both(\sigma_i^\both(z))=z$.
    While the \gls{BC} in \eqref{eq:sys_intermediate_bc0} remains unchanged under the spatial transformation \eqref{eq:ztrafo} in that
    \begin{equation}
        \bar x_+(0,t) = Q_0 \bar x_-(0,t),
    \end{equation}
    one has
    \begin{equation}
    \label{eq:sys_scaled_bc1}
        \bar x_{-,i}(\sigma_i^-(1),t) = \tilde u_i(t), \qquad i = 1,\dots,n_-
    \end{equation}
\end{subequations}
instead of \eqref{eq:sys_intermediate_bc1}.
In contrast to \eqref{eq:sys_intermediate}, the homogenized velocities in \eqref{eq:sys_scaled_pde} result in different spatial domains $\bar z\in[0,\sigma_i^\both(1)]$ for the components $\bar x_{\both,i}(\bar z,t)$, $i=1,\dots,n_\both$.

The reduced spatial domain, apparent from $\sigma_i^\both(1)\le1$ for $i=1,\dots,n_\both$, motivates the introduction of a dynamic extension that complements the spatial domain such that $\bar z\in[0,1]$.
Hence, the domain for each of the components $\bar w_{\both,i}(\bar z,t)$, $i=1,\dots,n_c^\both$, with $n_c^\both=n_\both-1$, of the controller state is chosen such that the composite state $\chi(z,t)\in\Rset^n$ of the dynamically extended plant is defined on $[0,1]$:
\begin{subequations}
\label{eq:state_ext}
    \begin{align}
        \chi_{\both,i}(z,t) &= \begin{cases}
        \bar x_{\both,i}(z,t), & z\in[0,\sigma_i^\both(1)] \\
        \bar w_{\both,i}(z,t), & z\in(\sigma_i^\both(1),1]
        \end{cases}
        \intertext{for $i=1,\dots,n_\both-1$ and}
        \chi_{\both,n_\both}(z,t) &= \bar x_{\both,n_\both}(z,t), \quad z\in[0,1].
    \end{align}
\end{subequations}
The controller dynamics
\begin{subequations}
\label{eq:dynext_pdes}
    \begin{align}
    \label{eq:dynext_pdes_m}
        \partial_t \bar w_{-,i}(\bar z,t) &= \phantom{-}\lambda_{n_-}^-(\bar z) \partial_{\bar z} \bar w_{-,i}(\bar z,t),
        \notag\\
        &\hspace{1cm} \bar z\in[\sigma_i^-(1),1),\,\,\,\, i = 1,\dots,n_c^- \\
    \label{eq:dynext_pdes_p}
        \partial_t \bar w_{+,i}(\bar z,t) &= -\lambda_{n_+}^+(\bar z) \partial_{\bar z} \bar w_{+,i}(\bar z,t),
        \notag\\
        &\hspace{1cm} \bar z\in(\sigma_i^+(1),1],\,\,\,\,\, i = 1,\dots,n_c^+
    \end{align}
\end{subequations}
reflect the same transport velocities as the spatially scaled plant PDEs in \eqref{eq:sys_scaled_pde}.
Note that \eqref{eq:dynext_pdes} is but one choice to achieve homogenization of transport velocities for the dynamically extended system.
One could at least consider the inclusion of additional local terms depending on $\bar x_-(0,t)$ in \eqref{eq:dynext_pdes}, analogously to \eqref{eq:sys_scaled_pde}.

In order to join the plant and controller PDEs (cf.\ \eqref{eq:sys_scaled_pde} and \eqref{eq:dynext_pdes}), the \glspl{BC} connecting both dynamics have to satisfy
\begin{subequations}
\label{eq:connect}
    \begin{align}
    \label{eq:connect_m}
        \bar x_{-,i}(\sigma_i^-(1),t) &= \bar w_{-,i}(\sigma_i^-(1),t), && i=1,\dots,n_c^-\\
    \label{eq:connect_p}
        \bar w_{+,i}(\sigma_i^+(1),t) &= \bar x_{+,i}(\sigma_i^+(1),t), && i=1,\dots,n_c^+.
    \end{align}
\end{subequations}
While \eqref{eq:connect_p} takes the role of the \gls{BC} of \eqref{eq:dynext_pdes_p}, \eqref{eq:connect_m} defines the input components.
Thus, based on \eqref{eq:sys_scaled_bc1}, the control input follows as
\begin{equation}
\label{eq:eq:connect_utilde}
    \tilde u_i(t) = \bar w_{-,i}(\sigma_i^-(1),t), \qquad i = 1,\dots,n_c^-,
\end{equation}
with $\tilde u_{n_-}(t)$ yet to be chosen.
The controller dynamics is completed by \glspl{BC}
\begin{equation}
\label{eq:new_input_1}
    \bar w_{-,i}(1,t) = v_i(t), \qquad i = 1,\dots,n_c^-
\end{equation}
for \eqref{eq:dynext_pdes_m}. Together with
\begin{equation}
\label{eq:new_input_2}
    v_{n_-}(t) = \tilde u_{n_-}(t),
\end{equation}
this yields the new input $v(t)\in\Rset^{n_-}$.

Based on \eqref{eq:state_ext}, the dynamically extended system consisting of the plant \eqref{eq:sys_scaled_pde} and the controller dynamics \eqref{eq:dynext_pdes}--\eqref{eq:new_input_2} reads
\begin{subequations}
\label{eq:sysext}
    \begin{align}
    \label{eq:sysext_pde}
        \partial_t \chi(z,t) &= \bar\Lambda(z) \partial_z \chi(z,t) + \bar A_0(z) \chi_-(0,t) \\
    \label{eq:sysext_bc0}
        \chi_+(0,t) &= Q_0 \chi_-(0,t) \\
    \label{eq:sysext_bc1}
        \chi_-(1,t) &= v(t),
  \end{align}
\end{subequations}
with $\chi(z,t)=\col(\chi_-(z,t),\chi_+(z,t))\in\Rset^n$, $(z,t)\in[0,1]\times\Rset^+$, homogenized transport velocities in
\begin{equation}
    \bar\Lambda(z) = \begin{bmatrix} \lambda_{n_-}^-(z)I_{n_-} & 0 \\ 0 & -\lambda_{n_+}^+(z)I_{n_+} \end{bmatrix}
\end{equation}
and a matrix $\bar A_0=\col(\bar A_0^-,\bar A_0^+)$ with piecewise continuous elements defined by
\begin{equation}
    e_i^\top \bar A_0^{\both}(z) =
    \begin{cases}
        e_i^\top A_0^{\both}(\tau_i^\both(z)), & z\in[0,\sigma_i^\both(1)] \\
        0
        , & z\in(\sigma_i^\both(1),1],
    \end{cases}
\end{equation}
$i=1,\dots,n_\both$.
Note that $\bar A_0^-(z)$ inherits its strictly lower triangular form from $A_0^-(z)$.
The \gls{IC} $\chi(z,0)=\chi_0(z)$ of \eqref{eq:sysext} follows from $x_0(z)$ and the \gls{IC} of the controller dynamics \eqref{eq:dynext_pdes} that can be chosen freely.

\subsection{Result of the dynamic extension}
\label{sec:ext_result}

The dynamic extension constitutes the controller dynamics and comprises \eqref{eq:dynext_pdes}, \eqref{eq:connect_p} and \eqref{eq:new_input_1}.
Based on the constructive derivation in the previous section, they can be rewritten in a condensed form analogous to that of the heterodirectional hyperbolic system \eqref{eq:sys}.
This is summarized in the following lemma.
For that, recall that the dimension $n_c^\both=n_\both-1$, with $n_c^-+n_c^+=n_c$, corresponds to the number of transport velocities of \eqref{eq:sys_pde} that are faster than the slowest one $\lambda_{n_\both}^\both(z)$ and define by $E_{c_+}\in\Rset^{n\times n_c^+}$ a matrix of the form $E_+$ in \eqref{eq:def_E}, with $n_c^+$ instead of $n_+$.


\begin{lem}[Dynamic extension]
\label{lem:extension}
    The dynamic extension \eqref{eq:dynext_pdes}, \eqref{eq:connect_p} and \eqref{eq:new_input_1} for the system \eqref{eq:sys} is given by the heterodirectional hyperbolic system
    \begin{subequations}
    \label{eq:ctrldyn}
        \begin{align}
        \label{eq:ctrldyn_pde}
            \partial_t w(z,t) &= \Gamma(z) \partial_z w(z,t) \\
        \label{eq:ctrldyn_bc0}
            w_+(0,t) &= E_{c_+}^\top \! \big[x_+(1,t) - \tint_0^1 K(1,\zeta) x(\zeta,t)\d\zeta\big] \\
        \label{eq:ctrldyn_bc1}
            w_-(1,t) &= \col(v_1(t),\dots,v_{n_c^-}(t))
        \end{align}
        of transport equations defined for $(z,t)\in[0,1]\times\Rset^+$, with the control input given by
        \begin{equation}
        \label{eq:u_final}
            u(t) = \tint_0^1 E_-^\top K(1,\zeta) x(\zeta,t)\d\zeta - Q_1 x_+(1,t) + \begin{bmatrix} w_-(0,t) \\ v_{n_-}(t) \end{bmatrix}\!.
        \end{equation}
    \end{subequations}
    The \gls{IC} $w(z,0)=w_0(z)$ for the controller state $w(z,t)=\col(w_-(z,t),w_+(z,t))\in\Rset^{n_c}$, $n_c=n-2$, is an arbitrary piecewise continuous function and the elements
    \begin{equation}
    \label{eq:map_z_ctrldyn}
        \gamma_i^\both(z) = \frac{\lambda_{n_\both}^\both(z+(1-z)\sigma_i^\both(1))}{1-\sigma_i^\both(1)}, \qquad i = 1,\dots,n_c^\both
    \end{equation}
    of $\Gamma = \diag(\Gamma^-,-\Gamma^+)$ with $\Gamma^\both = \diag(\gamma_1^\both,\dots,\gamma_{n_c^\both}^\both)$ are $C^1([0,1])$-functions that satisfy $0 < \gamma_1^\both(z) < \cdots < \gamma_{n_c^\both}^\both(z)$, $z\in[0,1]$.
\end{lem}

\begin{pf}
    The map $z = \rho_i^\both(\bar z) = (\bar z-\sigma_i^\both(1))/(1-\sigma_i^\both(1))$ is used to scale the spatial domain $[\sigma_i^\both(1),1]$ of \eqref{eq:dynext_pdes} to the interval $[0,1]$ considered in \eqref{eq:ctrldyn_pde}.
    Then, introducing
    \begin{equation}
    \label{eq:wbar}
        w_{\both,i}(z,t) = w_{\both,i}(\rho_i^\both(\bar z),t) = \bar w_{\both,i}(\bar z,t), \quad i = 1,\dots,n_c^\both,
    \end{equation}\eqref{eq:dynext_pdes} gives \eqref{eq:ctrldyn_pde} with the transport velocities \eqref{eq:map_z_ctrldyn}.
    Moreover, combining \eqref{eq:wbar}, \eqref{eq:connect_p} and \eqref{eq:xbar} gives $w_{+,i}(0,t) = \tilde x_{+,i}(1,t)$, $i=1,\dots,n_c^+$, from which \eqref{eq:ctrldyn_bc0} follows in view of the backstepping transformation \eqref{eq:trafo1}.
    The \gls{BC} \eqref{eq:ctrldyn_bc1} is easily verified based on \eqref{eq:new_input_1}.
    Finally, \eqref{eq:utilde} together with \eqref{eq:eq:connect_utilde} and \eqref{eq:new_input_2} yields \eqref{eq:u_final}.
    \hfill $\Box$
\end{pf}

Note that the assignment of the control input $u(t)$ in \eqref{eq:u_final} takes the role of an output w.r.t.\ the controller dynamics, with the new (delayed) control input $v(t)\in\Rset^{n_-}$.
It should also be pointed out that one could choose more complex controller dynamics beyond pure transport equations as in \eqref{eq:ctrldyn_pde}, e.g., ones that contain additional local terms depending on $x_-(0,t)$ analogous to \eqref{eq:sys_intermediate}.
In addition, one could introduce $\Gamma(z)$ such that the dynamically extended system \eqref{eq:sysext} admits constant transport velocities.
Based on the specific dynamic extension chosen here, the following result is obtained.

\begin{thm}[Dynamically extended system]
    Applying the dynamic extension \eqref{eq:ctrldyn} with the controller state $w(z,t)\in\Rset^{n_c}$ to the system \eqref{eq:sys} with the state $x(z,t)\in\Rset^n$ and the input $u(t)\in\Rset^{n_-}$ yields a dynamically extended system that can be represented by \eqref{eq:sysext} with the state $\chi(z,t)\in\Rset^n$ and the input $v(t)\in\Rset^{n_-}$.
\end{thm}

The proof directly follows from the constructive derivation in Section~\ref{sec:ext_derivation} and the invertible backstepping transformation \eqref{eq:trafo1}.

\section{Dynamic state feedback controllers}
\label{sec:applications}

The newly introduced dynamically extended system \eqref{eq:sysext} offers to accommodate various design specifications beyond what can be accomplished by a static feedback of the state $x(z,t)$ of \eqref{eq:sys}.
In what follows, dynamic state feedback controllers for $v(t)$ are designed to meet different objectives by appropriately choosing a static feedback of the extended state $\chi(z,t)$ of \eqref{eq:sysext} and combining it with the controller dynamics \eqref{eq:ctrldyn}.
First, as an alternative to the design in \cite{Redaud2022CDC,Redaud2024SCL}, a closed-loop dynamics with arbitrary in-domain couplings is assigned using backstepping.
Then, dynamic controllers are shown to allow a decoupling between the components of the input and some given output.
Finally, the concept of dynamic extension is directly applied to hyperbolic PDE-ODE systems, thus emphasizing the potential of the dynamic state feedback design presented here.

\subsection{General closed-loop dynamics}
\label{sec:arbdynamics}

Based on the dynamically extended system \eqref{eq:sysext}, the backstepping approach can be used to assign general closed-loop dynamics
\begin{subequations}
\label{eq:systarget}
    \begin{align}
    \label{eq:systarget_pde}
        \partial_t \bar\chi(z,t) &= \bar\Lambda(z) \partial_z \bar\chi(z,t) + \bar B(z) \bar\chi(z,t) \\
    \label{eq:systarget_bc0}
        \bar\chi_+(0,t) &= Q_0 \bar\chi_-(0,t) \\
    \label{eq:systarget_bc1}
        \bar\chi_-(1,t) &= B_0 \bar\chi_-(0,t) + B_1 \bar\chi_+(1,t) + \! \tint_0^1 \! B(\zeta) \bar\chi(\zeta,t)\d \zeta,
  \end{align}
\end{subequations}
with $(z,t)\in[0,1]\times\Rset^+$ and $\bar\chi(z,t)=\col(\bar\chi_-(z,t),\allowbreak \bar\chi_+(z,t))\in\Rset^n$.
Therein, the matrices $B_0$, $B_1$, $B(z)$ and $\bar B(z)$ with $b_{ij},\bar b_{ij}\in C([0,1])$ are freely assignable.
In comparison to the usual approach that uses static state feedback, this makes possible parametrizing a much larger class of target systems.

In order to map the dynamically extended dynamics \eqref{eq:sysext} into the target system \eqref{eq:systarget}, consider the (inverse) Volterra integral transformation
\begin{align}
\label{eq:trafo2}
    \chi(z,t) &= M(z)\bar\chi(z,t) + \tint_0^z L(z,\zeta) \bar\chi(\zeta,t) \d\zeta
\end{align}
(inspired by \cite[Rem.~6]{Hu2016tac}) with matrices $M(z)\in\Rset^{n\times n}$, $z\in[0,1]$, and $L(z,\zeta)\in\Rset^{n\times n}$, $(z,\zeta)\in\mathcal T$, of the form
\begin{equation}
\label{eq:defMK}
    M = \begin{bmatrix} M^- & 0 \\ 0 & M^+ \end{bmatrix}, \qquad
    L = \begin{bmatrix} L^{--} & L^{-+} \\ L^{+-} & L^{++} \end{bmatrix}
\end{equation}
that are partitioned according to the states $\chi_-(z,t)$ and $\chi_+(z,t)$.
Using the same subdivision for the coupling matrix $\bar B$ as for the kernel $L$, it is found that $M(z)$ has to solve the initial value problem
\begin{subequations}
\label{eq:ivp}
    \begin{align}
    \label{eq:ivp_m}
        \d_z M^-(z) &= \phantom{-}\tfrac{1}{{\lambda_{n_-}^-(z)}} M^-(z) \bar B^{--}(z) \\
    \label{eq:ivp_p}
        \d_z M^+(z) &= -\tfrac{1}{\lambda_{n_+}^+(z)} M^+(z) \bar B^{++}(z) \\
        M(0) &= I_n
    \end{align}
\end{subequations}
for $z\in[0,1]$ and $L(z,\zeta)$ has to solve the kernel equations%
\begin{subequations}
\label{eq:kernel2}
    \begin{align}
    \label{eq:kernel2_pde}
        & \bar\Lambda(z)\partial_z L(z,\zeta) + \partial_\zeta(L(z,\zeta)\bar\Lambda(\zeta)) = L(z,\zeta)\bar B(\zeta) \\
    \label{eq:kernel2_mp}
        & L^{-+}(z,z) = \phantom{-}\frac{1}{\lambda_{n_-}^-(z)+\lambda_{n_+}^+(z)} M^-(z) \bar B^{-+}(z) \\
    \label{eq:kernel2_pm}
        & L^{+-}(z,z) = -\frac{1}{\lambda_{n_-}^-(z)+\lambda_{n_+}^+(z)} M^+(z) \bar B^{+-}(z) \\
    \label{eq:kernel2_mm}
        & \lambda_{n_+}^+(0) L^{-+}(z,0)Q_0 - \lambda_{n_-}^-(0) L^{--}(z,0) = \bar A_0^-(z) \!\! \\
    \label{eq:kernel2_pp}
        & \lambda_{n_+}^+(0) L^{++}(z,0)Q_0 - \lambda_{n_-}^-(0) L^{+-}(z,0) = \bar A_0^+(z) \!\!
    \end{align}
\end{subequations}
on $\mathcal T$.
A comparison of \eqref{eq:systarget_bc1} and \eqref{eq:sysext_bc1} yields the input%
\begin{multline}
\label{eq:feedback_v}
    v(t) = M^-(1)\big[B_0 \bar\chi_-(0,t) + B_1 \bar\chi_+(1,t)\big] \\
     + \tint_0^1 \big[M^-(1)B(\zeta) + E_-^\top L(1,\zeta)\big] \bar\chi(\zeta,t)\d\zeta.
\end{multline}
In view of $\lambda_{n_\both}^\both\in C^1([0,1])$ and $\bar b_{ij}\in C([0,1])$, it is directly implied by \cite{Kailath1980} that the initial value problem \eqref{eq:ivp} has a unique solution $M(z)\in\Rset^{n\times n}$, with $\det M(z)\neq0$ for $z\in[0,1]$ and $m_{ij}\in C^1([0,1])$.
In the following lemma, it is verified that \eqref{eq:kernel2} admits a piecewise continuous solution if $\rank Q_0=n_-$, thus implying $n_+\ge n_-$, with uniqueness ensured by an additional artificial \gls{BC}.

\begin{lem}[Kernel equations]
\label{lem:kernel}
    If $\rank Q_0=n_-$, then there exists a matrix $R_0\in\Rset^{n_+\times(n_+-n_-)}$ with $\det[Q_0,R_0]\neq0$ and an artificial \gls{BC}
    \begin{equation}
    \label{eq:kernel2_artificalBC}
        L^{++}(z,0)R_0 = R(z),
    \end{equation}
    where the elements of $R(z)\in\Rset^{n_+\times(n_+-n_-)}$ are arbitrary, piecewise continuous functions, such that the kernel equations \eqref{eq:kernel2} admit a unique piecewise continuous solution $L(z,\zeta)$.
\end{lem}

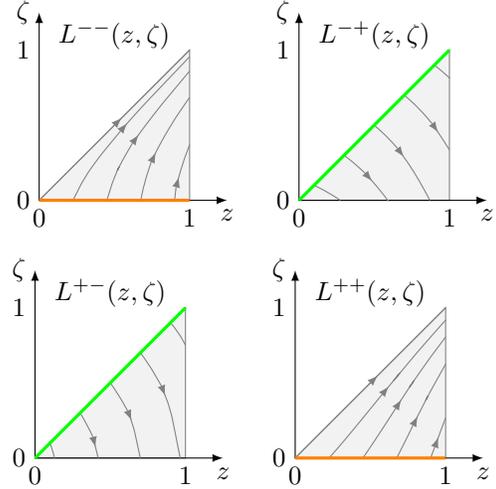
\begin{figure}
    \centering
    \begin{tikzpicture}
    \tikzlot{$L^{--}(z,\zeta)$}
    \draw[gray] (0,0) -- (1,1);
    \draw[gray,latex reversed-] (1,1) -- (2,2);
    \draw[gray,rounded corners] (0.45,0) -- (0.6,0.3) -- (0.8,0.58) -- (1.08,0.92);
    \draw[gray,rounded corners,latex reversed-] (1.08,0.92) -- (1.2,1.06) -- (1.4,1.28) -- (1.6,1.49) -- (1.8,1.7) -- (2,1.91);
    \draw[rounded corners,gray] (0.9,0) -- (1,0.32) -- (1.1,0.52) -- (1.24,0.76);
    \draw[rounded corners,gray,latex reversed-] (1.24,0.76) -- (1.4,0.98) -- (1.6,1.24) -- (1.8,1.48) -- (2,1.72);
    \draw[rounded corners,gray] (1.35,0) -- (1.4,0.24) -- (1.49,0.51);
    \draw[rounded corners,gray,latex reversed-] (1.49,0.51) -- (1.6,0.74) -- (1.8,1.08) -- (2,1.37);
    \draw[gray,rounded corners] (1.8,0) -- (1.82,0.175);
    \draw[gray,rounded corners,latex reversed-] (1.82,0.175) -- (1.9,0.47) -- (2,0.74);
    \draw[very thick, orange] (0,0) -- (2,0);
\end{tikzpicture} \,\,
\begin{tikzpicture}
    \tikzlot{$L^{-+}(z,\zeta)$}
    \draw[gray,rounded corners] (0.2,0.2) -- (0.4,0.1) -- (0.55,0);
    \draw[gray,rounded corners] (0.6,0.6) -- (0.8,0.44) -- (0.88,0.36);
    \draw[gray,rounded corners,latex reversed-] (0.88,0.36) -- (1.1,0.11) -- (1.18,0);
    \draw[gray,rounded corners] (1,1) -- (1.2,0.82) -- (1.31,0.69);
    \draw[gray,rounded corners,latex reversed-] (1.31,0.69) -- (1.45,0.50) -- (1.6,0.26) -- (1.74,0);
    \draw[gray,rounded corners] (1.4,1.4) -- (1.6,1.22) -- (1.77,1.02);
    \draw[gray,rounded corners,latex reversed-] (1.77,1.02) -- (1.9,0.83) -- (2,0.67);
    \draw[gray,rounded corners] (1.8,1.8) -- (1.9,1.72) -- (2,1.63);
    \draw[very thick, green] (0,0) -- (2,2);
\end{tikzpicture} \\[1ex]
\begin{tikzpicture}
    \tikzlot{$L^{+-}(z,\zeta)$}
    \draw[gray,rounded corners=.7ex] (0.2,0.2) -- (0.23,0.12) --  (0.26,0);
    \draw[gray,rounded corners] (0.6,0.6) -- (0.7,0.43) -- (0.77,0.28);
    \draw[gray,rounded corners,latex reversed-] (0.77,0.28) -- (0.84,0);
    \draw[gray,rounded corners] (1,1) -- (1.1,0.86) -- (1.23,0.63);
    \draw[gray,rounded corners,latex reversed-] (1.23,0.63) -- (1.3,0.46) -- (1.4,0);
    \draw[gray,rounded corners] (1.4,1.4) -- (1.5,1.27) -- (1.68,0.96);
    \draw[gray,rounded corners,latex reversed-] (1.68,0.96) -- (1.8,0.65) -- (1.93,0);
    \draw[gray,rounded corners] (1.8,1.8) -- (1.9,1.69) -- (2,1.5);
    \draw[very thick, green] (0,0) -- (2,2);
\end{tikzpicture} \,\,
\begin{tikzpicture}
    \tikzlot{$L^{++}(z,\zeta)$}
    \draw[gray] (0,0) -- (1,1);
    \draw[gray,latex reversed-] (1,1) -- (2,2);
    \draw[gray,rounded corners] (0.45,0) -- (0.6,0.17) -- (0.8,0.41) -- (1,0.655) -- (1.155,0.845);
    \draw[gray,rounded corners,latex reversed-] (1.155,0.845) -- (1.3,1.02) -- (1.4,1.14) -- (1.6,1.38) -- (1.8,1.61) -- (2,1.836);
    \draw[rounded corners,gray] (0.9,0) -- (1,0.145) -- (1.2,0.45) -- (1.336,0.66);
    \draw[rounded corners,gray,latex reversed-] (1.336,0.66) -- (1.5,0.914) -- (1.6,1.06) -- (1.8,1.345) -- (2,1.61);
    \draw[rounded corners,gray] (1.35,0) -- (1.5,0.298) -- (1.566,0.434);
    \draw[rounded corners,gray,latex reversed-] (1.566,0.434) -- (1.7,0.70) -- (1.8,0.893) -- (2,1.25);
    \draw[gray,rounded corners] (1.8,0) -- (1.855,0.145);
    \draw[gray,rounded corners,latex reversed-] (1.855,0.145) -- (1.93,0.35) -- (2,0.535);
    \draw[very thick, orange] (0,0) -- (2,0);
\end{tikzpicture}
    \caption{Typical characteristics visualize the kernel equations \eqref{eq:kernel2}, with the \glspl{BC} at $\zeta=0$ and $\zeta=z$ highlighted in color. Due to the common transport velocities in $\bar\Lambda^-(z)$ and $\bar\Lambda^+(z)$, respectively, each block of $L(z,\zeta)$ in \eqref{eq:defMK} is represented by only one element.}
    \label{fig:kernel}
\end{figure}

This result follows as a special case of \cite{Hu2019siam}.
By the homogenized transport velocities in $\bar\Lambda(z)$, the kernel equations \eqref{eq:kernel2} can easily be solved similar to the scalar case where $n_-=n_+=1$ (see, e.g., \cite{Vazquez2011CDC,Coron2013SIAM}).
Due to $\rank Q_0=n_-$ and $\det[Q_0,R_0]\neq0$, one can uniquely solve \eqref{eq:kernel2_pp} and \eqref{eq:kernel2_artificalBC} for $L^{++}(z,0)$.
Together with \eqref{eq:kernel2_mp}--\eqref{eq:kernel2_mm}, this specifies the boundary values\footnote{Contrary to the case of distinct transport velocities in \cite{Hu2019siam}, this already ensures well-posedness of the kernel equations without the need for additional artificial \glspl{BC}.} for each of the respective four block matrices of $L(z,\zeta)$ as illustrated in Figure \ref{fig:kernel}.
Based on that, the method of characteristics allows to rewrite \eqref{eq:kernel2_pde} as integral equations depending on the elements of the piecewise continuous elements of $\bar B(z)$.
Using the method of successive approximations and showing the convergence of the resulting series along the lines of \cite{Hu2016tac,Hu2019siam} verifies a unique piecewise $C(\mathcal T)$-solution $L(z,\zeta)$ for \eqref{eq:kernel2}.

\begin{rem} 
\label{rem:Q0}
    The assumption $\rank Q_0=n_-$ in Lemma~\ref{lem:kernel} can only be met if $n_+\ge n_-$. 
    However, this holds without loss of generality, since a further dynamic extension can always be introduced if $\rank Q_0=n_{q_0}<n_-$. 
    For that, augment the controller dynamics by $n_--n_{q_0}$ additional transport equations $\partial_t \omega_{+,i}(z,t) = -\lambda_{n_+}^+(z) \partial_z \omega_{+,i}(z,t)$, $z\in(0,1]$, $i = 1,\dots,n_--n_{q_0}$, where the \gls{BC} $\omega_+(0,t)=P_0\chi_-(0,t)$ is chosen such that $\rank\bar Q_0=n_-$, with $\bar Q_0=\col(Q_0,P_0)\in\Rset^{(n-n_{q_0})\times n_-}$. Then, $\chi_+(z,t)\mapsto\col(\chi_+(z,t),\omega_+(z,t))$ in the extended system \eqref{eq:sysext} yields $\chi_+(0,t)=\bar Q_0\chi_-(0,t)$.
    \hfill $\triangle$
\end{rem}

The bounded invertibility of the Volterra integral transformation \eqref{eq:trafo2}, where the kernel $L_\mathrm{I}(z,\zeta)\in\Rset^{n\times n}$ of the inverse transformation
\begin{equation}
\label{eq:trafo2_inv}
    \bar\chi(z,t) = M^{-1}(z) \chi(z,t) - \tint_0^z L_\mathrm{I}(z,\zeta)\chi(\zeta,t)\d\zeta
\end{equation}
follows from the reciprocity relation $L_\mathrm{I}(z,\zeta) = M^{-1}(z)L(z,\zeta) M^{-1}(\zeta) - \tint_\zeta^z M^{-1}(z)L(z,\bar\zeta) L_\mathrm{I}(\bar\zeta,\zeta) \d\bar\zeta$, together with the definition of $\chi(z,t)$ in \eqref{eq:state_ext} and the preliminary transformation \eqref{eq:trafo1} ensures that \eqref{eq:feedback_v} can be rewritten as a feedback of the system state $x(z,t)$ and the controller state $w(z,t)$.
Stability of \eqref{eq:systarget} is ensured by appropriate choice of the design matrices $B_0$, $B_1$, $B(z)$ and $\bar B(z)$.
In the most trivial case, setting all of these matrices equal to zero guarantees finite-time stability (see Remark~\ref{rem:cl_stability}).
However, more generally, the design matrices allow for choosing physically motivated target systems \eqref{eq:systarget} (see \cite{Redaud2022CDC,Redaud2024SCL}).
For simplicity, the following theorem only addresses the case of asymptotic stability of the closed loop.

\begin{thm}[Closed-loop stability]
\label{thm}
    Let $B_0$, $B_1$, $B(z)$ and $\bar B(z)$ be such that \eqref{eq:systarget} is asymptotically stable pointwise in space and assume $\rank Q_0=n_-$.
    Then, the closed-loop system that results from applying the dynamic state feedback \eqref{eq:ctrldyn} and \eqref{eq:feedback_v} to the system \eqref{eq:sys} is asymptotically stable pointwise in space for all piecewise continuous \glspl{IC} $x(z,0) = x_0(z)\in\Rset^n$ and $w(z,0) = w_0(z)\in\Rset^{n_c}$.
\end{thm}

\begin{pf}
    By an appropriate choice of $B_0$, $B_1$, $B(z)$ and $\bar B(z)$, asymptotic stability of \eqref{eq:systarget} pointwise in space means that $\bar\chi(z,t)\to0$ for $t\to\infty$, $z\in[0,1]$ and any piecewise continuous \gls{IC} $\bar\chi(z,0) = \bar{\chi}_0(z) \in \mathbb R^n$.
    The latter is obtained from the piecewise continuous functions $x_0$ and $w_0$ based on \eqref{eq:trafo1}, \eqref{eq:xbar}, \eqref{eq:wbar}, \eqref{eq:state_ext} and \eqref{eq:trafo2_inv}.
    Then, by \eqref{eq:trafo2}, $\chi(z,t)\to0$ for $t\to\infty$.
    The same asymptotic convergence holds for $\bar x(\bar z,t)$ and $\bar w(\bar z,t)$ in light of \eqref{eq:state_ext} as well as for the states $\tilde x(z,t)$ and $w(z,t)$ on the normalized spatial domain with $z\in[0,1]$ (see \eqref{eq:xbar} and \eqref{eq:wbar}).
    Finally, the bounded invertibility of \eqref{eq:trafo1}, similar to \eqref{eq:trafo2_inv}, verifies that $x(z,t)\to0$ for $t\to\infty$ and $z\in[0,1]$.
    \hfill $\Box$
\end{pf}



\begin{rem}
\label{rem:cl_stability}
    It is well known that choosing $B_0$, $B_1$, $B(z)$ and $\bar B(z)$ equal to zero ensures finite-time stability of \eqref{eq:systarget} in minimum time $T_\mathrm{min}=\phi_{n_-}^-(1)+\phi_{n_+}^+(1)$ (see \cite{Auriol2016AUT,Coron2017AUT}), i.e., $\bar\chi(z,t)=0$ for $t>T_\mathrm{min}$, which implies $x(z,t)=0$ for $t>T_\mathrm{min}$.
    Note that $T_\mathrm{min}$ corresponds to the sum of the delays induced by the slowest transport velocities $\lambda_{n_-}^-(z)$ and $\lambda_{n_+}^+(z)$ of system \eqref{eq:sys} in the negative and positive $z$-direction, respectively.
    As such, the dynamic extension for homogenization of transport velocities and the use of the dynamic feedback \eqref{eq:ctrldyn} and \eqref{eq:feedback_v} do not increase the theoretical lower bound for the control time of \eqref{eq:sys}.
    \hfill $\triangle$
\end{rem}

\subsection{Decoupling control}
\label{sec:decoupling}

It is well known that non-interacting control of finite-dimensional systems requires dynamic controllers, in general (see \cite{Isidori1995} for nonlinear systems).
Though the concept of decoupling control for PDE systems is not new (see, e.g., \cite{Curtain1985} for a different class of systems), here, it arises very naturally.
Consider, for example, $y(t) = x_-(0,t)$ as the output of system \eqref{eq:sys}.
While finding a feedback for $u(t)$ such that \eqref{eq:sys} admits a decoupled closed-loop transfer behavior w.r.t.\ a new input or doing so for $\tilde u(t)$ based on \eqref{eq:sys_intermediate} is anything but obvious, decoupling between $x_-(0,t)=\chi_-(0,t)$ and a new input $\bar v(t)$ to be introduced is easily achieved for the dynamically extended system \eqref{eq:sysext}.
In fact, one only has to remove the local terms associated with $\bar A_0^-(z)$ in the PDE \eqref{eq:sysext_pde} to obtain the completely decoupled input-output behavior $y(t)=\bar v(t-\phi_{n_-}^-(1))$ for arbitrary $\bar v(t)$.
For that, consider a special case
\begin{multline}
\label{eq:trafo2_decoupling}
    \chi_-(z,t) = \bar\chi_-(z,t) + \tint_0^z \big[L^{--}(z,\zeta) \bar\chi_-(\zeta,t) \\
    + L^{-+}(z,\zeta) \chi_+(\zeta,t)\big] \d\zeta
\end{multline}
of the Volterra integral transformation \eqref{eq:trafo2}, where $\chi_+(z,t)$ is left unchanged.
Consequently, if $L^{--}(z,\zeta)$ and $L^{-+}(z,\zeta)$ as introduced in \eqref{eq:defMK} satisfy the kernel equations \eqref{eq:kernel2_pde}, \eqref{eq:kernel2_mp} and \eqref{eq:kernel2_mm} with $\bar B(z)=0$, then \eqref{eq:trafo2_decoupling} maps \eqref{eq:sysext} into the form
\begin{subequations}
\label{eq:sysdecoupl}
    \begin{align}
        \partial_t \bar\chi_-(z,t) &= \phantom{-}\lambda_{n_-}^-(z) \partial_z \bar\chi_-(z,t) \\
        \partial_t \chi_+(z,t) &= -\lambda_{n_+}^+(z) \partial_z \chi_+(z,t) \!+\! \bar A_0^+(z) \bar\chi_-(0,t) \!\! \\
        \chi_+(0,t) &= Q_0 \bar\chi_-(0,t) \\
    \label{eq:sysdecoupl_1}
        \bar\chi_-(1,t) &= v(t) - \tint_0^1 \big[L^{--}(1,\zeta) \bar\chi_-(\zeta,t) \\
        &\hspace{1.5cm} + L^{-+}(1,\zeta) \chi_+(\zeta,t)\big] \d\zeta \eqqcolon \bar v(t). \nonumber
    \end{align}
\end{subequations}
The desired input-output decoupling is achieved by introducing the new input $\bar v(t)$ in \eqref{eq:sysdecoupl_1}.
Again, this feedback is easily rewritten in terms of the system state $x(z,t)$ and the controller state $w(z,t)$ using \eqref{eq:trafo2_inv}, \eqref{eq:state_ext} and \eqref{eq:trafo1}.

In fact, system \eqref{eq:sysdecoupl} resolves into two parts, the input-output decoupled dynamics
\begin{subequations}
\label{eq:decoupl_io}
    \begin{align}
        \partial_t \bar\chi_-(z,t) &= \lambda_{n_-}^-(z) \partial_z \bar\chi_-(z,t) \\
        \bar\chi_-(1,t) &= \bar v(t) \\
        y(t) &= \bar\chi_-(0,t)
    \end{align}
\end{subequations}
and the internal dynamics
\begin{subequations}
\label{eq:decoupl_internal}
    \begin{align}
        \partial_t \chi_+(z,t) &= -\lambda_{n_+}^+(z) \partial_z \chi_+(z,t) + \bar A_0^+(z) \bar\chi_-(0,t) \\
        \chi_+(0,t) &= Q_0 \bar\chi_-(0,t).
    \end{align}
\end{subequations}
The latter is input-to-state stable w.r.t.\ the input $\bar\chi_-(0,t)$.
For $\bar\chi_-(0,t)=0$, it is also well-known as the zero dynamics (e.g., \cite{Jacob2019AUT}), which is finite-time stable here.
Note the close relation to zero dynamics for finite-dimensional systems as well as the Byrnes-Isidori normal form (see \cite{Isidori1995} for nonlinear systems and \cite{Ilchmann2016SIAM} for a class of PDE systems).

\subsection{Dynamic state feedback for PDE-ODE systems}
\label{sec:pdeode}

The idea of homogenization as well as its applications in Sections \ref{sec:arbdynamics} and \ref{sec:decoupling} for the assignment of general closed-loop dynamics and decoupling control, respectively, can easily be transferred to heterodirectional hyperbolic PDEs with an additional ODE dynamics at the unactuated boundary.
To briefly sketch this, consider the bidirectionally coupled PDE-ODE system
\begin{subequations}
\label{eq:PDEODE}
    \begin{align}
    \label{eq:PDEODE_ode}
        \dot\xi(t) &= F \xi(t) + B x_-(0,t) \\
    \label{eq:PDEODE_0}
        x_+(0,t) &= Q_0 x_-(0,t) + C_0 \xi(t) \\
    \label{eq:PDEODE_pde}
        \partial_t x(z,t) &= \Lambda(z) \partial_z x(z,t) + A(z) x(z,t) + C(z) \xi(t) \\
    \label{eq:PDEODE_1}
        x_-(1,t) &= Q_1 x_+(1,t) + u(t),
    \end{align}
\end{subequations}
with ODE state $\xi(t)\in\Rset^{n_0}$ and PDE state $x(z,t)=\col(x_-(z,t),x_+(z,t))\in\Rset^n$.
While the boundary value $x_-(0,t)$ drives the ODE, with the pair $(F,B)$ assumed to be at least stabilizable, the ODE state acts on the PDE subsystem \eqref{eq:PDEODE_0}--\eqref{eq:PDEODE_1} via $C_0\in\Rset^{n_+\times n_0}$ and $C(z)\in\Rset^{n\times n_0}$ with $C([0,1])$ elements.
All remaining quantities are defined identically to those of the heterodirectional hyperbolic system \eqref{eq:sys} in Section~\ref{sec:problem}.

While there are several strategies to design a (static) backstepping-based state feedback for \eqref{eq:PDEODE} (e.g., \cite{DiMeglio2018aut}), here, the design in \cite{Deutscher2019IJC} is combined with the dynamic extension for homogenization in Section~\ref{sec:homogenization}.
Specifically, in \cite{Deutscher2019IJC}, the PDE-ODE system \eqref{eq:PDEODE} is mapped into the form
\begin{subequations}
\label{eq:PDEODE_target}
    \begin{align}
    \label{eq:PDEODE_target_ode}
        \dot\xi(t) &= (F+BK) \xi(t) + B \varepsilon_-(0,t) \\
    \label{eq:PDEODE_target_0}
        \varepsilon_+(0,t) &= Q_0 \varepsilon_-(0,t) \\
    \label{eq:PDEODE_target_pde}
        \partial_t \varepsilon(z,t) &= \Lambda(z) \partial_z \varepsilon(z,t) + A_0(z) \varepsilon_-(0,t) \\
    \label{eq:PDEODE_target_1}
        \varepsilon_-(1,t) &= \tilde u(t),
    \end{align}
\end{subequations}
with $K$ such that $F+BK$ is Hurwitz and
\begin{multline}
    \tilde u(t) = Q_1 x_+(1,t) + u(t) - \tint_0^1 E_-^\top K(1,\zeta)x(\zeta,t)\d\zeta \\
    - \tint_0^1 E_-^\top P_\mathrm{I}(1,\zeta)\varepsilon(\zeta,t)\d\zeta - E_-^\top N_\mathrm{I}(1)\xi(t),
\end{multline}
by combining the preliminary Volterra integral transformation \eqref{eq:trafo1} that removes the in-domain coupling due to $A(z)$ and a second transformation
\begin{align}
    \tilde x(z,t) &= \varepsilon(z,t) + \tint_0^z P_\mathrm{I}(z,\zeta) \varepsilon(\zeta,t) \d\zeta + N_\mathrm{I}(z) \xi(t).
\end{align}
The latter stabilizes the ODE subsystem \eqref{eq:PDEODE_ode} and cascades the transport equations of the PDE subsystem \eqref{eq:PDEODE_target_0}--\eqref{eq:PDEODE_target_1} (see \cite{Deutscher2019IJC} for the kernel equations and their solution).
Based on \eqref{eq:PDEODE_target}, the results of Section~\ref{sec:homogenization} can be directly applied in order to homogenize the transport velocities by dynamic extension.
While the dynamic extension could have already been introduced following the preliminary transformation \eqref{eq:trafo1}, this is not done here for simplicity of presentation, i.e., to directly combine the results in \cite{Deutscher2019IJC} with those in Section~\ref{sec:homogenization}.
In the end, assigning a general closed-loop dynamics (similar to Section~\ref{sec:arbdynamics}) or decoupling the closed-loop transfer behavior according to Section \ref{sec:decoupling} is mostly straightforward.
Note that the closed-loop dynamics of the PDE subsystem (cf.~\eqref{eq:systarget}) could contain additional dependencies on the ODE state.
Also, instead of selecting $y(t)=x_-(0,t)\in\Rset^{n_-}$, it is interesting to consider an output $y(t)=H\xi(t)\in\Rset^{n_-}$, with $H\in\Rset^{n_-\times n_0}$.


\section{Example}
\label{sec:example}

Consider a system \eqref{eq:sys} with matrices $\Lambda^-(z) = \Lambda^+(z) = \text{diag}(2+\frac{1}{2}z,1+z)$,
\begin{equation}
    A(z) = (1-z)^2\begin{bmatrix} 0 & z & z^3 & -\frac{1}{2}-z \\ -z & 0 & \frac{1}{4}-z & -2+z \\ -z^3 & -\frac{1}{4}+z & 0 & \tan\frac{z}{3} \\ \frac{1}{2}+z & 2-z & -\tan\frac{z}{3} & 0 \end{bmatrix}
\end{equation}
and $Q_0 = Q_1 = -I_2$, where $n_+=n_-=2$.
The output $y(t)=x_-(0,t)\in\Rset^2$ of the system is the boundary value distal to the control input $u(t)\in\Rset^2$.
In order to ensure a complete input-output decoupling of the closed-loop transfer behaviors in Section~\ref{sec:decoupling}, a dynamic extension is introduced that homogenizes the transport velocities.
It can be written as the transport equations \eqref{eq:dynext_pdes} on the spatial domain $[\frac{9}{16},1]$ following from $\sigma_1^-(z)=\sigma_1^+(z)=\frac{1}{2}z+\frac{1}{16}z^2$ (cf.\ \eqref{eq:ztrafo}) or in the form \eqref{eq:ctrldyn} with $n_c^-=n_c^+=1$ and $\gamma^-(z)=\gamma^+(z)=\frac{25}{7}+z$ (see \eqref{eq:map_z_ctrldyn}).
Then, choosing $v(t)$ as in \eqref{eq:sysdecoupl_1} introduces the new input $\bar v(t)$ that ensures the completely decoupled input-output behavior $y(t) = \bar v(t-\phi_2^-(1))$.
Thereby, internal stability of the decoupled closed-loop system is guaranteed by the input-to-state stable internal dynamics \eqref{eq:decoupl_internal}.

\begin{figure}
    \centering
    \includegraphics[width=\columnwidth]{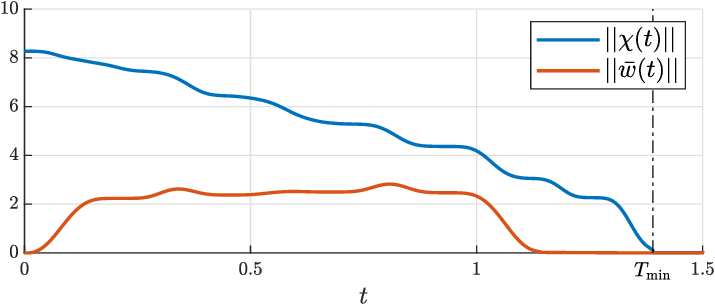}
    \caption{Norms $||\chi(t)||$ and $||\bar w(t)||$ of the dynamically extended system state and the controller state in closed loop.}
    \label{fig:norm}
\end{figure}

\begin{figure}
    \centering
    \includegraphics[width=\columnwidth]{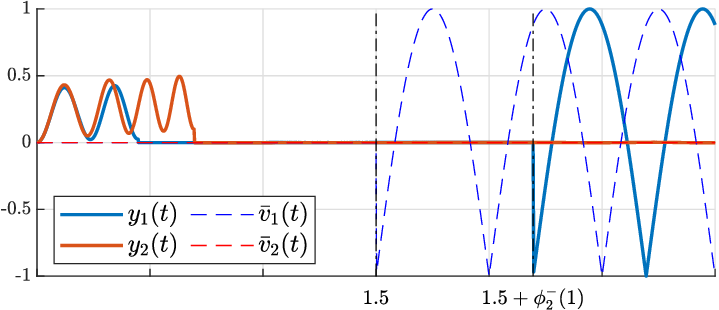}
    \includegraphics[width=\columnwidth]{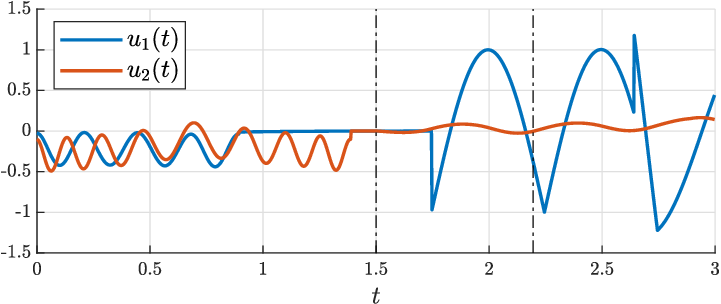}
    \caption{Evolution of $y(t)$, $\bar v(t)$ and $u(t)$.}
    \label{fig:decoupling}
\end{figure}

Using \glspl{IC} $x_-(z,0)=[\frac{2}{5}\sin^2(2\pi z),\frac{2}{5}\sin^2(4\pi z)]^\top=-x_+(z,0)$ and $w(z,0)=0$, $z\in[0,1]$, the results in Figures \ref{fig:norm} and \ref{fig:decoupling} are obtained by spatially discretizing the system and controller along the characteristic curves and applying the Euler method for numerical integration.
For the new input, $\bar v_1(t) = (|2\sin(2\pi t)|-1) h(t-1.5)$ with the Heaviside function $h(t)$ and $\bar v_2(t)=0$ are chosen in order to highlight two key aspects: convergence to zero in minimum time $T_\text{min}=\phi_{n_-}^-(1)+\phi_{n_+}^+(1)\approx1.39$ as $\bar v(t)=0$ for $t<1.5$ (e.g., \cite{Auriol2016AUT}) and decoupling between the input $\bar v(t)$ and the output $y(t)$ in the closed-loop system.
Specifically, in Figure~\ref{fig:norm}, the norm $||\chi(t)||$ of the extended system state that comprises the states of system and controller confirms $x(z,t)=0$, $z\in[0,1]$, after the minimum time $T_\text{min}$, in spite of the dynamic extension essentially delaying the control action.
Furthermore, as $y(t)=\bar v(t-\phi_2^-(1))$ for $t\ge\phi_2^-(1)$, this means that $y_1(t)$ follows the reference specified by $\bar v_1(t)$ while $y_2(t)$ is held at zero (see Figure~\ref{fig:decoupling}).
The decoupling of $\bar v(t)$ and $y(t)$ is further emphasized by the fact that both components of $u(t)$ are non-zero for $t\ge1.5$.

\section{Concluding remarks}


The presented design of dynamic state feedback controllers makes use of a dynamic extension that homogenizes the transport velocities.
Then, a static feedback of the extended state allows to meet different design objectives.
Importantly, it is not possible to achieve input-output decoupling in the closed loop or a target system with arbitrary in-domain couplings by a static feedback of the system state (see \cite{Redaud2022CDC,Redaud2024SCL} for the latter).
In particular, this means that the local terms introduced in the target system by a classical Volterra integral transformation for the well-posedness of the kernel equations are a consequence of using only a static state feedback.
The look at PDE-ODE systems also showed the great potential of the systematic introduction of dynamic extensions, with a variety of possible generalizations to different classes of distributed-parameter systems, including parabolic PDEs.
Future work will also be concerned with gaining a better understanding of different types of state feedback controllers for distributed-parameter systems, especially as they relate to endogenous and exogenous dynamic state feedback (see, e.g., \cite{Levine2009} in the finite-dimensional case).
Among other things, this is an important basis for generalizing the flatness-based control design in \cite{Woittennek2013CPDE,Gehring2023at} to the multivariable case (recall the dynamic feedback for the drilling example in \cite{Knueppel_2014_chapter}).

\begin{ack}
    This research was funded in part by the Austrian Science Fund (FWF) under project no.\ I 6519-N and the Deutsche Forschungsgemeinschaft (DFG, German Research Foundation) under project no.\ 517291864.
    For the purpose of Open Access, the author has applied a CC BY public copyright licence to any Author Accepted Manuscript (AAM) version arising from this submission.
    The authors also want to thank Frank Woittennek from UMIT Tirol, Austria, for his valuable suggestions.
\end{ack}

\appendix

\bibliographystyle{plain}        
\bibliography{mybib}           

\end{document}